%% file: Generalization_uncertain_MIMO_dynamic.tex
\documentclass{ifacconf}

\usepackage{amsmath,amssymb,here}
\usepackage{bm}

\newtheorem{hyp}{\bf Assumption}[section]

\newtheorem{lemma}{\bf Lemma}[section]

\newtheorem{proposition}{\bf Proposition}[section]

\usepackage{tikz}
\usetikzlibrary{arrows,shapes,calc}
\usepackage{pgfplots}
\usetikzlibrary{positioning} 

\usepackage{graphicx}      
\usepackage{natbib}        
\begin{document}
\begin{frontmatter}

\title{Output Feedback Control of a Class of Uncertain Systems Under Control-derivative dependent disturbances} 

\author[First]{M. Alamir}
\author[second]{J. Dobrowolski}
\author[First]{A. T. Mohammed} 

\address[First]{CNRS, University of Grenoble Alpes (mazen.alamir(amgad.mohamed)@gipsa-lab.grenoble-inp.fr).}
\address[second]{Schneider Electric Industrie. 37 Quai Paul Louis Merlin. 38000 Grenoble (jean.dobrowolski@schneider-electric.com)}

\begin{abstract}                
This paper addresses the design of robust dynamic output feedback control for highly uncertain systems in which the unknown disturbance might be excited by the derivative of the control input. This context appears in many industrial problems such as the speed control of the hydraulic turbines and the frequency stabilization in micro grids to cite but few examples. A key feature that has to be carefully addressed in this context is that too agressive feedback might lead to the loss of controllability and/or a significant drop in the closed-loop performance. The paper formulates the problem, underlines its relevance and gives a rigorous solution in which a dynamic output feedback is given together with a realistic set of sufficient conditions on the controller's parameters that enable to assess the behavior of the closed-loop under different circumstances. A numerical example is given to illustrate the relevance of the proposed successful design characterization.  
\end{abstract}


\end{frontmatter}
\section{Introduction}
This paper addresses the problem of output feedback design of dynamical systems showing the following features:
\begin{enumerate}
\item There is a simple although uncertain relationship between $u$ and $y$ ($y$ is of relative degree $0$). This simplicity suggests the use of appealingly simple feedback design.\\
\item However, a cautionless design of such a simple law might excite some internal dynamics of the system leading to oscillations and/or lack of controllability.\\
\item In spite of this issue, the use of {\em advanced} observer-based nonlinear controller that would be designed using the entire model in order to master the internal dynamics is not welcomed by practitioners because of its high complexity and the need to derive a tight models of complex behaviors that are not always well mastered. Moreover, the observer design is not always easy nor the system is always observable. 
\end{enumerate} 
These reasons make the framework addressed in the present contribution appealing (when possible), namely, keep simple control structure but set the parameters of the simple control law in a sound and theoretically assessed way so that the above mentioned coupling issue is appropriately addressed. \ \\ \ \\ To the best of the authors knowledge, existing works do not tackle the case of output feedback of uncertain systems with control derivative-dependent uncertainties. Typical classes of uncertainties that are commonly handled are parametric uncertainties \citep{KAR1999}, additive bounded state-dependent uncertainties \citep{Chen1998,Nunes2009} or nonlinear uncertainties with affine state dependent bounds \citep{li2016adaptive} to cite but few representative references. \ \\ \ \\ Works belonging to the model-free approach \citep{Fliess2013}, while addressing higher relative degrees, are also based on a representation of the unknown term to be estimated that is not dependent on the control behavior in a way that might enhance instable coupling. \ \\ \ \\ 
This paper is organized as follows: First, the problem is stated in section \ref{secps} and motivated through the description of two industrial example (control of hydraulic turbines and frequency stabilization in micro grids).  Section \ref{secwa} discusses the working assumptions. The main results regarding the successful feasibility bounds are given in Section \ref{seccond}. An illustrative example is given in Section \ref{secex}. Finally, Section \ref{secconc} concludes the paper and gives some hints for further investigation.\ \\ \ \\  
In the sequel, given a time interval $I:=[t_1,t_2]$ and a variable $x$, the boldfaced symbol $\bm x$ is used to denote the profile $\bigl\{x(t)\bigr\}_{t\in I}$ of $x$ over $I$ when the later is known without ambiguity from the context. Otherwise the explicit notation $\bm x^{(I)}$ is used instead. Let $\bm x_t:=x(t)$ and if $x$ is scalar, let $\|\bm x\|_\infty:=\sup_{t\in I}\vert x(t)\vert$. The set of possible profiles $\bm w$ of the exogenous uncertainty  involved in (\ref{syst2}) is denoted by $\mathbb W$. \section{Problem Statement} \label{secps} 
We consider uncertain dynamic systems of the form:
\begin{align}
\dot y(t)&=\alpha(t)\Bigl[u(t)-g(t)+\ell(t,\eta(t)))\Bigr] \label{syst1}\\
\dot\eta(t)&=E(\bm\eta^{[t-\tau,t]},\dot{\bm u}^{[t-\tau,t]},\bm w^{[t-\tau,t]}) \label{syst2} 
\end{align} 
where $y\in \mathbb{R}$ and $u\in [\underline u,\overline u]$ are the regulated output and the control input respectively. $g(\cdot)$ is an unknown term satisfying $g(t)\in [\underline g,\overline g]$ with known bounds $\underline g$ and $\overline g$. The vector $\eta\in \mathbb{R}^{n_\eta}$ represents internal states with dynamics given by (\ref{syst2}) in which $E: \mathbb{R}^{n_\eta}\times \mathbb{R}\times \mathbb{R}^{n_w}\rightarrow \mathbb{R}^{n_\eta}$ is possibly unknown map. the vector $w\in \mathbb{R}^{n_w}$ represents exogenous disturbances. Finally $\alpha(t)\in [\underline \alpha,\bar\alpha]$ is an unknown scalar that might be time-varying, with known bounds $\underline\alpha$ and $\bar\alpha$. Let $\Delta_u:=\bar u-\underline u$ and $\Delta_g=\bar g-\underline g$ denotes the admissible excursion of the control input and the unknown term $g$ respectively. 
\\
\begin{rem}
Equation (\ref{syst2}) is a rather general representation that can encompasses time delayed systems, general partial differential equations to cite but few examples. 
\end{rem}

A careful examination of (\ref{syst1}) clearly shows that the controllability of $y$ can be guaranteed if one can enforce the following inclusion:
\begin{equation}
h:=g-\ell(\eta)\in [\underline u+\rho,\overline u-\rho]\subset [\underline u,\bar u] \label{enforce} 
\end{equation} 
for some $\rho>0$ and if $\vert\dot u\vert$ can be taken arbitrarily high since in this case, $u$ can {\em instantaneously} dominate $h$ and consequently enforce any desired sign to $\dot y$. However, the fact that the dynamic (\ref{syst2}) of $\eta$  depends on $\dot u$ makes this second condition unrealistic as a strong dynamics on $u$ (high $\dot u$) can steer the system to regions where (\ref{enforce}) is violated leading to the loss of controllability and even instability. This is precisely the issue that is investigated in this contribution. Note that the absence of precise knowledge of $\ell(\cdot,\cdot)$, $E(\cdot,\cdot,\cdot)$, $g(\cdot)$ and $\alpha$ makes the problem even more challenging as it prevents, for instance, the reconstruction of $\eta$ (from the only measured $y$) and the use of the estimated values in the necessarily careful control design. Moreover, the observability itself may not hold in many situations. \ \\ \ \\ 
In this paper, the following simple dynamic output feedback is considered 
\begin{align}
u&=S\bigl(\lambda(y_d-y)+z\bigr) \label{defdeu}\\
\dot z&=\ \lambda_f(u-z) \label{defdezdot} 
\end{align} 
where $S$ is the saturation function that projects the argument inside $[\underline u,\overline u]$ and $y_d$ is the constant set-point for the regulated output $y$. This feedback is defined up to the choice of the two positive parameters $\lambda_f,\lambda>0$. The problem addressed in this paper can be formulated as follows: \ \\ \ \\ 
\begin{minipage}{0.1\textwidth}
\sc Problem
\end{minipage} 
\begin{minipage}{0.02\textwidth}
\rule{0.1mm}{23mm} 
\end{minipage} 
\begin{minipage}{0.33\textwidth}
\sl Find a high level characterization of the uncertainties and maps involved in (\ref{syst1})-(\ref{syst2}) and associated conditions on the control parameters $(\lambda,\lambda_f)$ so that the behaviour of the closed-loop system can be conveniently/quantifiably assessed.
\end{minipage} 
\vskip 0.2cm 
More precisely, the assessment of the closed-loop properties should be understood in the following sense:
\begin{enumerate}
\item Asymptotic stability should be obtained in the absence of disturbance ($w\equiv 0$, $g\equiv 0$ and $\ell\equiv 0$)\\
\item A quantifiable bound can be given on the asymptotic tracking error in the case where persistent dynamic $g$ and non vanishing disturbance $w$ are present. 
\end{enumerate} 
Before addressing the problem stated above, the following section underlines the relevance of the framework described by (\ref{syst1})-(\ref{syst2}) by showing that it describes at least two concrete industrial problems. 
\subsection{Relevance of the framwork} \label{secrelev} 
The framework described above fits at least two important industrial problems that are briefly discussed here.\subsubsection{Control of hydraulic turbines}
\ \\ \ \\ 
Efficient control of hydraulic turbines in the context of Power Storage Plant (PSP) is gaining a renewed interest. In this context, the hydraulic plant is used as a mean to compensate for the intermittency of renewable sources of energy \citep{connolly2012}. As such, the response time of speed control is a key parameter that determines the success of the whole power management system. In this context, the regulated variable $y$ is the rotational speed $\Omega$, the control $u$ is the guiding vane's opening $\gamma$ while the internal state $\eta$ is the state of the penstock (flow rate and pressure values inside the penstock) that drives the water flow from the water reservoir to the turbine's level. All practitioners know \citep{Zhou2011} that the major obstacle that limits high rotational acceleration of the turbine lies in the induced oscillating pressure waves in the penstock. When referring to the internal dynamic (\ref{syst2}), the time laps $\tau$ represent twice the time needed for the pressure wave to travel the penstock's length. These pressure oscillations generate disturbing torque variations at the turbine's level that can go beyond what the controlled torque can compensate if the dynamics of the control is not monitored appropriately. That is the reason why simple control laws that are not carefully tuned against this phenomenon fail to produce satisfactory results. 
\subsubsection{Distributed frequency-control in isolated micro grids}
\ \\ \ \\ 
This is another example that falls into the framework studied in this paper. Here the regulated variable is the grid frequency $f$. The control input $u$ is the sum of produced powers by the Gensets and the renewable sources while the uncertainty $w$ is the derivative of the load power consumption. The internal dynamics is induced by the flexibility of power transmission between Gensets following sudden change in the load power demand. The need for a distributed control design (in which only the frequency measurement is used) emphasizes the need for simple control laws that cannot afford the task of model-based reconstruction of the internal dynamic (which needs both observability and total knowledge of the model's details). This internal dynamics disturbs the commonly used \citep{Zhao2016,Xialing2014} power balance equation $\dot f=\alpha(P-P_L)$ by an additional term such that:
\begin{align}
\dot f&=\alpha(u-P_L+C\eta) \label{systf1}\\
\dot\eta&=A\eta+B_1\dot u+B_2\dot P_L \label{systf2} 
\end{align} 
which obviously takes the form (\ref{syst1})-(\ref{syst2}) with $w=\dot P_L$. Note that a 4th order dynamics is typically needed to correctly represent the phenomena when only two Gensets are simply considered. 
\section{Working Assumptions} \label{secwa} 
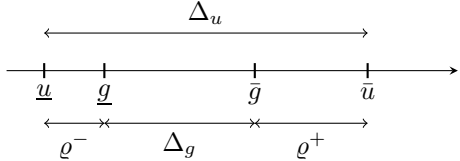
\begin{figure}[H]
\begin{center}
\begin{tikzpicture}
\draw[->,>=stealth] (0,0) -- (6,0);
\draw[thick] (0.5,-0.1) -- (0.5,+0.1) node[below=1.5mm]{$\footnotesize \underline u$};
\draw[thick] (1.3,-0.1) -- (1.3,+0.1) node[below=1.5mm]{$\footnotesize \underline g$};
\draw[thick] (3.3,-0.1) -- (3.3,+0.1) node[below=1.5mm]{$\footnotesize \bar g$};
\draw[thick] (4.8,-0.1) -- (4.8,+0.1) node[below=1.5mm]{$\footnotesize \bar u$};
\draw[<->] (0.5,0.5) -- node[midway,above]{$\footnotesize \Delta_u$} (4.8,0.5);
\draw[<->] (1.3,-0.7) -- node[midway,below]{$\footnotesize \Delta_g$} (3.3,-0.7);
\draw[<->] (0.5,-0.7) -- node[midway,below]{$\footnotesize \varrho^-$} (1.3,-0.7);
\draw[<->] (3.3,-0.7) -- node[midway,below]{$\footnotesize \varrho^+$} (4.8,-0.7);
\end{tikzpicture}
\end{center} 
\caption{Definition of the different positive constants involved in the characterization of the actuator and the unknown term $g$.} \label{defdesparamDelta} 
\end{figure}

In this section, the high level characterization that are mentioned in the problem's statement (Section \ref{secps}) are introduced. These characterizations delimit the class of unknown dynamics and quantities for which the simple law (\ref{defdeu})-(\ref{defdezdot}) can be appropriate provided that some conditions on the choice of $\lambda$ and $\lambda_f$ are fulfilled. These conditions are established in Section \ref{seccond}. \\ \ \\  
The first unavoidable assumption is called the {\em Control Authority Margin} (CAM) assumption. It simply states that $u$ can {\em ultimately} dominates the unknown term $g\in [\underline g,\bar g]$ alone with some margins, denoted by $\varrho^+>0$ and $\varrho^-$:
\begin{hyp}[Control Authority Margin]\label{hyp1} \ \\ 
There exist two scalars $\varrho^->0$ and $\varrho^+>0$ such that:
\begin{equation}
\bar u\ge \bar g+\varrho^+\quad \mbox{\rm and}\quad \underline u\le \underline g-\varrho^- 
\end{equation} 
These different positive constants are shown in Figure \ref{defdesparamDelta}. 
\end{hyp}
Note that without this assumption, there is obviously no means to guarantee bounded behavior of the output tracking error and this, regardless of the control law being used. 
The next assumption characterizes the behavior of the additional term $\ell$ involved in (\ref{syst1}) under the dynamic (\ref{syst2}) for all possible disturbance profiles $\bm w\in \mathbb W$ and all profiles $\bm u$:
\begin{hyp}\label{hyp2} 
There exist positive scalars $c_0, d_0>0$ and $c_1, d_1>0$ such that, for all $\bm w\in \mathbb{W}$ and all admissible input profiles $\bm u$ (inside $[\underline u,\bar u]$), the following inequality holds:
\begin{align}
\vert \ell(t,\eta(t))\vert &\le c_0+c_1\|\dot{\bm u}^{[t-\tau,t]}\|_\infty \label{eqass2} \\
\vert \dot\ell(t,\eta(t))\vert &\le d_0+d_1\|\dot{\bm u}^{[t-\tau,t]}\|_\infty \label{eqass3} 
\end{align} 
Moreover, the evolution of the state $\eta$ remains bounded for any bounded profiles in  $\|\bm u\|_\infty$ and $\|\dot{\bm u}\|_\infty$.
\end{hyp}
This assumption can be better understood if one considers the case where the dynamics (\ref{syst2}) is given by (\ref{systf2}) in which the matrix $A$ is Hurwitz (open-loop stable internal dynamics). Indeed in this case, which holds for both industrial contexts described in Section \ref{secrelev}, the inequalities  (\ref{eqass2})-(\ref{eqass3}) become true after a finite time\footnote{beyond which the influence of the initial state that prevailed at the beginning of the system's first operation time becomes lower than $\epsilon$.} with:
\begin{align}
c_0&:= \epsilon+\sup_{\bm w\in \mathbb W}\sup_{t\ge 0}\left\vert C\int_0^te^{A(t-\sigma)}B_2\bm w_\sigma d\sigma\right\vert  \label{defdec0lin}\\
c_1&:=\sup_{t\ge 0}\left\vert C\int_0^te^{A(t-\sigma)}B_1d\sigma\right\vert \label{defdec1lin}\\ 
d_0&:= \epsilon+\sup_{\bm w\in \mathbb W}\sup_{t\ge 0}\left\vert CA\int_0^te^{A(t-\sigma)}B_2\bm w_\sigma d\sigma\right\vert  \label{defded0lin}\\
d_1&:=\vert CB_1\vert+\sup_{t\ge 0}\left\vert CA\int_0^te^{A(t-\sigma)}B_1d\sigma\right\vert \label{defdec1lin}
\end{align} 
In the case of hydraulic turbines, asymptotic stability of the penstock dynamics comes from the fact that this system is theoretically marginally stable but becomes asymptotically stable by introducing a pre-compensator in the definition of the flow rate feedback. The control $u$ discussed in this paper represent then the definition of the remaining feed-froward term.  As for the micro grids case, the asymptotic stability comes from the use of the droop frequency control as a first rapid loop on the top of which, the control we are interested in amounts at translating the droop curve vertically. This is obviously equivalent to an action on the total produced power $u$. \ \\ \ \\ 
In the sequel, the short notation $\|\dot{\bm u}\|_\infty$ is used for $\|\dot{\bm u}^{[t-\tau,t]}\|_\infty$ as $t$ is removed and $\tau$ is supposed to be fixed once for all. 
\section{Sufficient Conditions on Control parameters and associated bounds on the tracking error} \label{seccond} 
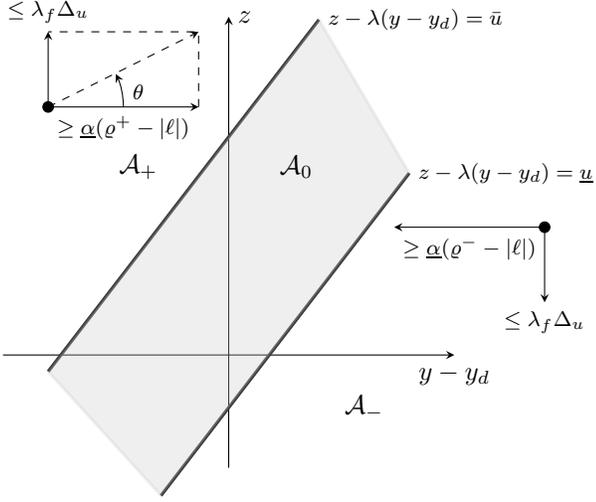
\begin{figure}
\begin{center}
\begin{tikzpicture}
\def\Lx{3}
\def\Ly{4}
\def\lam{1.3}
\def\umin{-0.3*\Ly}
\def\umax{0.6*\Ly}
\def\xm{-0.3*\Lx}
\def\zm{0.8*\Lx}
\def\ym{\lam*\xm+\umin}
\def\qm{\lam*\zm+\umin}
\def\xi{-0.8*\Lx}
\def\zi{0.4*\Lx}
\def\yi{\lam*\xi+\umax}
\def\qi{\lam*\zi+\umax}
\draw[->,>=stealth] (-\Lx,-0.5) -- (\Lx,-0.5) node[below]{$y-y_d$}; 
\draw[->,>=stealth] (0,-0.5*\Ly) -- (0,\Ly) node[right]{$z$}; 
\draw[very thick,black,domain=-0.3*\Lx:0.8*\Lx,smooth,variable=\x] plot ({\x},{\lam*\x+\umin});
\draw[very thick,black,domain=-0.8*\Lx:0.4*\Lx,smooth,variable=\x] plot ({\x},{\lam*\x+\umax});
\node at(\xm,\ym) (A1){};
\node at(\zm,\qm) (A2){};
\node at(\xi,\yi) (A4){};
\node at(\zi,\qi) (A3){};
\draw[fill,very thick,black!20,opacity=0.3] (A1.base) -- (A2.base) -- (A3.base) -- (A4.base) -- cycle;
\node[right] at(A3) {\footnotesize $z-\lambda(y-y_d)=\bar u$};
\node[right] at(A2) {\footnotesize $z-\lambda(y-y_d)=\underline u$};
\node[black] at(0.3*\Lx,0.5*\Ly){$\mathcal A_0$};
\node[black] at(-0.4*\Lx,0.5*\Ly){$\mathcal A_+$};
\node[black] at(0.6*\Lx,-0.3*\Ly){$\mathcal A_-$};
\node at(-0.8*\Lx,0.7*\Ly) (O1){};
\draw[->,>=stealth] (O1.base) -- ($(O1)+(2,0)$) node[below,midway]{\footnotesize $\ge \underline\alpha(\varrho^+-\vert\ell\vert)$};
\draw[->,>=stealth] (O1.base) -- ($(O1)+(0,1)$) node[above]{\footnotesize $\le \lambda_f\Delta_u$};
\draw[dashed,thin] ($(O1)+(2,0)$) |- ($(O1)+(0,1)$);
\draw[dashed,thin,->,>=stealth] (O1.base) -- ($(O1)+(2,1)$);
\draw[->,>=stealth] ($(O1)+(1,0)$) arc (0:26:1);
\node at($(O1)+(1.2,0.2)$) {\footnotesize $\theta$};
\draw[fill=black] (O1) circle (2pt) ;
\node at(1.4*\Lx,0.3*\Ly) (O1){};
\draw[->,>=stealth] (O1.base) -- ($(O1)+(-2,0)$) node[below,midway]{\footnotesize $\ge \underline\alpha(\varrho^--\vert \ell\vert)$};
\draw[->,>=stealth] (O1.base) -- ($(O1)+(0,-1)$) node[below]{\footnotesize $\le \lambda_f\Delta_u$};
\draw[fill=black] (O1) circle (2pt) ;
\end{tikzpicture}
\end{center} 
\caption{Definition of the sets $\mathcal A_+$, $\mathcal A_-$ and $\mathcal A_0$ and disposition of the vector fields.} \label{thesets} 
\end{figure}

Let us first of all examine some properties of the signals involved in the feedback laws (\ref{defdeu})-(\ref{defdezdot}). 
\begin{lemma}\label{lem1} 
Under the feedback law (\ref{defdeu})-(\ref{defdezdot}), for all $\lambda>0$ satisfying:
\begin{equation}
\lambda<\dfrac{1}{\bar\alpha c_1} \label{condlam1} 
\end{equation} 
the following inequality holds:
\begin{equation}
\|\dot{\bm u}\|_\infty\le \delta_u:=\dfrac{\lambda\bar\alpha\bigl[\beta+c_0\bigr]+\lambda_f\Delta_u}{1-\lambda\bar\alpha c_1}  \label{tocheck} 
\end{equation} 
where
\begin{equation}
\beta:=\Delta_g+\max\bigl\{\varrho^+,\varrho^-\bigr\} \label{defdebeta} 
\end{equation} 
\end{lemma}
{\sc Proof}. By the very definition (\ref{defdeu}), inequality (\ref{tocheck}) has to be checked only inside the admissible domain as otherwise $\dot u=0$ and the inequality obviously holds. Now, when the saturation constraint is not active, one obviously has according to (\ref{defdeu}):
\begin{equation}
\vert \dot u\vert \le \lambda\vert \dot y\vert +\vert \dot z\vert \label{proof11} 
\end{equation} 
but according to (\ref{syst1}), it comes that:
\begin{equation}
\vert \dot y\vert \le \bar\alpha\Bigl[\vert u-g\vert +\vert\ell\vert\Bigr] \label{dotyhgy} 
\end{equation} 
an examination of Figure \ref{defdesparamDelta} clearly shows that:
\begin{equation}
\vert u-g\vert \le \beta:=\Delta_g+\max\Bigl\{\varrho^+,\varrho^-\Bigr\} \label{gftrgjk0} 
\end{equation}  
therefore, using (\ref{gftrgjk0}) and (\ref{eqass2}) in (\ref{dotyhgy}) gives:
\begin{equation}
\vert \dot y\vert \le \bar\alpha\Bigl[\beta+c_0+c_1\|\dot{\bm u}\|_\infty\Bigr] \label{gfre43} 
\end{equation} 
on the other hand, since $z$ is a first order filtered version of $u$, it comes that $z\in [\underline u,\bar u]$ and hence:
\begin{equation}
\vert \dot z\vert \le \lambda_f \Delta_u \label{kjhhu55} 
\end{equation} 
using  (\ref{gftrgjk0})-(\ref{gfre43}) in (\ref{proof11}) leads to:
\begin{equation}
\vert \dot u\vert \le \lambda \Bigl[\bar\alpha\beta+c_0+c_1\|\dot{\bm u}\|_\infty\Bigr]+\lambda_f\Delta_u
\end{equation} 
which can be rewritten equivalently as follows:
\begin{equation}
(1-\lambda\bar\alpha c_1)\|\dot{\bm u}\|_\infty\le \lambda\bar\alpha(\beta+c_0)+\lambda_f\Delta_u
\end{equation} 
which gives obviously (\ref{tocheck}) if (\ref{condlam1}) is satisfied. $\hfill \Box$ \ \\ \ \\
In order to go further in the analysis of the closed-loop behavior, the following sets are defined (see Figure \ref{thesets}):
\begin{align}
\mathcal A_+&:=\left\{(y,z)\ \vert \ z-\lambda(y-y_d)\ge \bar u\right\} \label{defdeAcalplus}\\ 
\mathcal A_-&:=\left\{(y,z)\ \vert \ z-\lambda(y-y_d)\le \underline u\right\} \label{defdeAcalmoins}\\
\mathcal A_0&:=\left\{(y,z)\ \vert \ z-\lambda(y-y_d)\in (\underline u,\bar u)\right\} \label{defdeAcal0}
\end{align}  
Note that by definition, if $(y,z)\in \mathcal A_+$ then $u=\bar u$, if $(y,z)\in \mathcal A_-$ then $u=\underline u$ while if $(y,z)\in \mathcal A_0$ then the control is not saturated, namely $u=\lambda(y_d-y)+z$.
\ \\ \ \\ 
Based on these definitions, the following result can be stated:
\begin{lemma}[$\mathcal A_0$ is attractive and invariant]\label{lemmelkjlk} \ \\ \ \\ 
{\bf If} the following conditions hold:
\begin{itemize}
\item $\lambda<1/(\bar\alpha c_1)$
\item $\lambda_f$ and $\lambda$ are such that 
\end{itemize}  
\begin{equation}
\lambda_f<\lambda\varphi(\lambda) \label{condonlambdaf} 
\end{equation}
where:
\begin{equation}
\varphi(\lambda):=\Bigl[\Delta_u\bigl(1+\dfrac{c_1\underline\alpha}{1-\lambda\bar\alpha c_1}\bigr)\Bigr]^{-1}\Bigl[\underline\varrho-c_0-\lambda\dfrac{(\beta+c_0)c_1\bar\alpha}{1-\lambda\bar\alpha c_1}\Bigr]\underline\alpha\label{LLA} 
\end{equation} 
where $\underline \varrho:=\min\{\varrho^-,\varrho^+\}$, 
{\bf then} the set $\mathcal A_0$ is attractive and invariant for the closed-loop dynamics associated to the control law (\ref{defdeu})-(\ref{defdezdot}) and to any constant desired value $y_d$. 
\end{lemma}
{\sc Proof}. The steps of the proof are the following:
\begin{enumerate}
\item any dynamics starting in $\mathcal A_+$ enters $\mathcal A_0$ in finite time. 
\item any dynamics starting in $\mathcal A_-$ enters $\mathcal A_0$ in finite time. 
\item $\mathcal A_0$ is invariant.
\end{enumerate}  
{\bf proof of (1)}. Assume that $(y,z)\in \mathcal A_+$ (see Figure \ref{thesets}. Examination of the vector field (see Figure \ref{thesets}) suggests that the result would be proved if one can prove that the following inequality holds:
\begin{equation}
\vert \dot z\vert<\lambda \vert\dot y\vert\label{cond111} 
\end{equation} 
On the other hand we know that $\vert \dot z\vert \le \lambda_f\Delta_u$ while $\vert \dot y\vert \ge \underline\alpha(\varrho_+-\vert\ell\vert)$. Now using the inequality (\ref{eqass2}), the inequality (\ref{cond111}) would be satisfied if the following condition holds:
\begin{equation}
\lambda_f\Delta_u<\underline\alpha\left[\varrho^+-(c_0+c_1\|\dot{\bm u}\|_\infty)\right]\lambda
\end{equation} 
and since $\lambda<1/(\bar\alpha c_1)$ the inequality (\ref{tocheck}) of Lemma \ref{lem1} holds, therefore,  (\ref{cond111}) would be satisfied if the following inequality holds:
\begin{equation}
\lambda_f\Delta_u<\underline\alpha\left[\varrho^+-\left[c_0+c_1\dfrac{\lambda\bar\alpha\bigl[\beta+c_0\bigr]+\lambda_f\Delta_u}{1-\lambda\bar\alpha c_1}\right]\right]\lambda
\end{equation} 
which is equivalent to (after straightforward manipulations):
\begin{equation}
\lambda_f<\Bigl[\Delta_u\bigl(1+\dfrac{c_1\underline\alpha}{1-\lambda\alpha c_1}\bigr)\Bigr]^{-1}\Bigl[\varrho^+-c_0-c_1\dfrac{(\beta+c_0)\bar\alpha\lambda}{1-\lambda\bar\alpha c_1}\Bigr]\underline\alpha\lambda \label{hgjy08} 
\end{equation} 
{\bf proof of (2)}. Following exactly the same arguments, it can be shown that $\mathcal A_0$ is attractive for any initial condition in $\mathcal A_-$ provided that the following condition holds:
\begin{equation}
\lambda_f<\Bigl[\Delta_u\bigl(1+\dfrac{c_1\underline\alpha}{1-\lambda\alpha c_1}\bigr)\Bigr]^{-1}\Bigl[\varrho^--c_0-c_1\dfrac{(\beta+c_0)\bar\alpha\lambda}{1-\lambda\bar\alpha c_1}\Bigr]\underline\alpha\lambda \label{hgjy09} 
\end{equation} 
Combining the more restrictive of the two conditions (\ref{hgjy08}) and (\ref{hgjy09}) obviously leads to (\ref{LLA}). \ \\ \ \\ 
{\bf Proof of (3)}. This simply comes from the fact that the flows $\dot x$ and $\dot z$ are continuous (in $y$ and $z$) and it has just been shown that the vector fields on the intersection boundaries $\mathcal A_+\cap \mathcal A_0$ and $\mathcal A_-\cap \mathcal A_0$ does strictly enter inside the set $\mathcal A_0$. Therefore $\mathcal A_0$ is invariant. $\hfill \Box$ \ \\ \ \\ 
Now regarding the existence of $\lambda,\lambda_f>0$ that satisfy the conditions of Lemma \ref{lemmelkjlk}, one has the following result:
\begin{lemma}[Existence of $\lambda,\lambda_f$] \label{lem34} 
If the following condition holds:
\begin{equation}
c_0<\underline\varrho:=\min\{\varrho^-,\varrho^+\} \label{hgt6587} 
\end{equation} 
then there exists $\lambda^*>0$ such that for all $\lambda<\lambda^*$, there exists $\lambda_f>0$ such that the pair $(\lambda,\lambda_f)$ satisfies the requirements of Lemma \ref{lemmelkjlk}. More precisely, $\lambda^*$ is solution of:
\begin{equation}
\lambda^*=\sup\left\{\sigma\le \dfrac{1}{\bar\alpha c_1}\quad\vert\quad \inf_{\lambda\in (0,\sigma)}\varphi(\lambda)\ge 0\right\} \label{defdelambdastar} 
\end{equation} 
\end{lemma}
{\sc Proof}. This is because, thanks to (\ref{hgt6587}), $\varphi(\cdot)$ satisfies:
\begin{equation}
\varphi(0)=\dfrac{\min\{\varrho^-,\varrho^+\}-c_0}{\Delta_u(1+c_1)}>0
\end{equation} 
and since $\varphi$ is a continuous function of $\lambda$ inside $(0,\frac{1}{\bar\alpha c_1})$, there exists $\lambda^*\in (0,\frac{1}{\bar\alpha c_1})$ such that for all $\lambda\in (0,\lambda^*)$, $\varphi(\lambda)>0$ hence there exists $\lambda_f>0$ satisfying (\ref{LLA}). $\hfill\Box$\ \\ \ \\ 
Based on the result of Lemma \ref{lemmelkjlk}, it comes out that the asymptotic behavior of the closed-loop is only determined by the behavior of the system under the non-saturated control law $u=\lambda(y_d-y)+z$. But this feedback law leads to a dynamics (inside $\mathcal A_0$) that can be written in the following form:
\begin{equation}
\begin{bmatrix}
\dot y\cr \dot z
\end{bmatrix}= \begin{bmatrix}
-\alpha\lambda & \alpha\cr -\lambda_f\lambda & 0
\end{bmatrix} \begin{bmatrix}
y-y_d\cr z-h
\end{bmatrix} \label{eqinsideA0} 
\end{equation}  
where $h$ is the unknown dynamics given by:
\begin{equation}
h=g-\ell(\eta) \label{defdeh} 
\end{equation}
and using the following notation:
\begin{equation}
e_y:=y-y_d\quad;\quad e_z:=z-h\quad;\quad A_0:=\begin{bmatrix}
-\alpha\lambda & \alpha\cr -\lambda_f\lambda & 0
\end{bmatrix} \label{nnkj78} 
\end{equation}  
the dynamics (\ref{eqinsideA0}) becomes:
\begin{equation}
\begin{bmatrix}
\dot e_y\cr \dot e_z
\end{bmatrix}= A_0 \begin{bmatrix}
e_y\cr e_z
\end{bmatrix}-\begin{bmatrix}
0\cr \dot h
\end{bmatrix}  \label{eqinsideA0bis} 
\end{equation} 
Based on this observation, the main result of the paper can be derived:
\begin{proposition}[Main result] \ \\ \ \\  
{\bf If} the following conditions are satisfied:
\begin{enumerate}
\item Assumptions \ref{hyp1}-\ref{hyp2} hold,
\item $c_0<\min\{\varrho^-,\varrho^+\}$,
\item $\lambda\in (0,\lambda^*)$ and $\lambda_f<\lambda\varphi(\lambda)$ where $\lambda^*$ solves (\ref{defdelambdastar}) in which $\varphi$ is given by (\ref{LLA}).
\item the unknown term $g$ satisfies $\vert \dot g\vert \le \delta_g$
\end{enumerate}  
{\bf then} the tracking error $e_y$ satisfies the following inequality:
\begin{align}
\lim_{t\rightarrow\infty}\vert y(t)-y_d\vert \le \dfrac{\delta_g+d_0+d_1\delta_u}{\lambda\lambda_f} \label{defdelimey}
\end{align} 
where $\delta_u>0$ is the constant given by (\ref{tocheck}). \ \\ \ \\ 
{\sc Proof}.  Under the Assumption of the proposition, the conditions of Lemma \ref{lemmelkjlk} (by definition of $\lambda^*$) hold. Therefore, the closed-loop trajectories lie ultimately inside $\mathcal A_0$ and the dynamics (\ref{eqinsideA0bis}) persistently prevails after a finite time. This means that one has asymptotically:
\begin{equation}
\begin{bmatrix}
e_y(t)\cr e_z(t)
\end{bmatrix} \rightarrow \int_0^te^{A_0(t-\sigma)} \begin{bmatrix}
0\cr \dot h(\sigma)
\end{bmatrix}d\sigma
\end{equation}   
Focusing on $e_y$ and taking the worst case leads to:
\begin{equation}
\lim_{t\rightarrow \infty}\vert e_y(t)\vert \le \left\vert \int_0^tC_1e^{A_0(t-\sigma)}C_2d\sigma\right\vert \max_{\sigma\in [0,t]}\vert \dot h(\sigma)\vert 
\end{equation} 
where $C_1:=(1,0)$ and $C_2=(0,1)^T$. This gives, because $A_0$ is invertible [see (\ref{nnkj78})]: 
\begin{equation}
\lim_{t\rightarrow \infty}\vert e_y(t)\vert \le \left\vert C_1A_0^{-1}C_2\right\vert\max_{\sigma\in [0,t]}\vert \dot h(\sigma)\vert \label{asbound1} 
\end{equation}
Now by definition (\ref{defdeh}) of $h$, the fact that $\vert \dot g\vert \le \delta_g$ and (\ref{tocheck}) of Lemma \ref{lem1}:
\begin{equation}
\max_{\sigma\in [0,t]}\vert \dot h(\sigma)\vert \le \delta_g+d_0+d_1\delta_u \label{asbound2} 
\end{equation} 
Moreover, simple computations show that $C_1A_0^{-1}C_2=1/(\lambda\lambda_f)$ which obviously gives the result. $\hfill \Box$
\end{proposition}
\section{Illustrative example} \label{secex} 
\begin{figure}
\begin{center}
\input{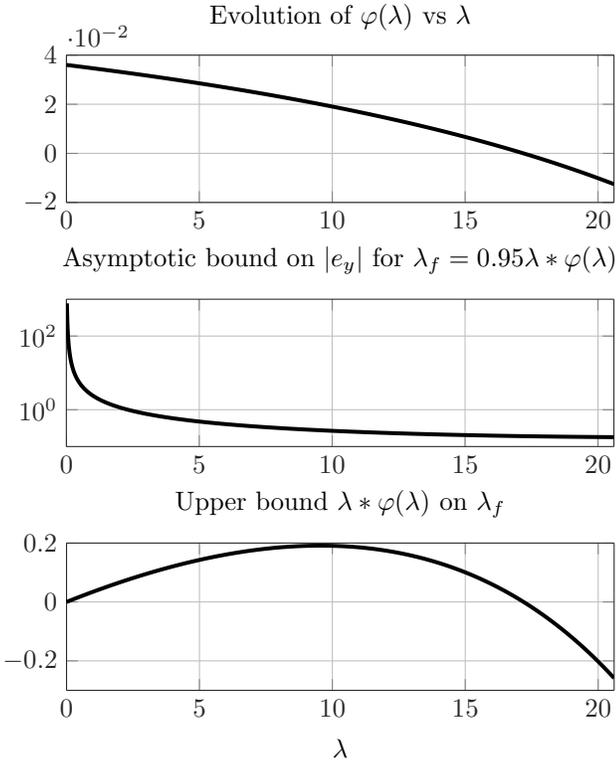}
\end{center} 
\caption{(top) Evolution of $\varphi(\lambda)$ given by (\ref{LLA}). (Middle) Asymptotic bound on the output tracking error $e_y$ [see (\ref{defdelimey})]. (Bottom) Upper bound $\lambda\varphi(\lambda)$ on the control parameter $\lambda_f$ [see (\ref{hgjy08})].} \label{analysis_plots} 

\end{figure}

\begin{figure}
\begin{center}
\input{figsimgood.tex}
\end{center} 
\caption{Behavior of the closed-loop system for the choice $(\lambda^{(1)},\lambda_f)=(17,0.0102)$ meeting the theoretical bounds.} \label{figsimgood} 
\end{figure}
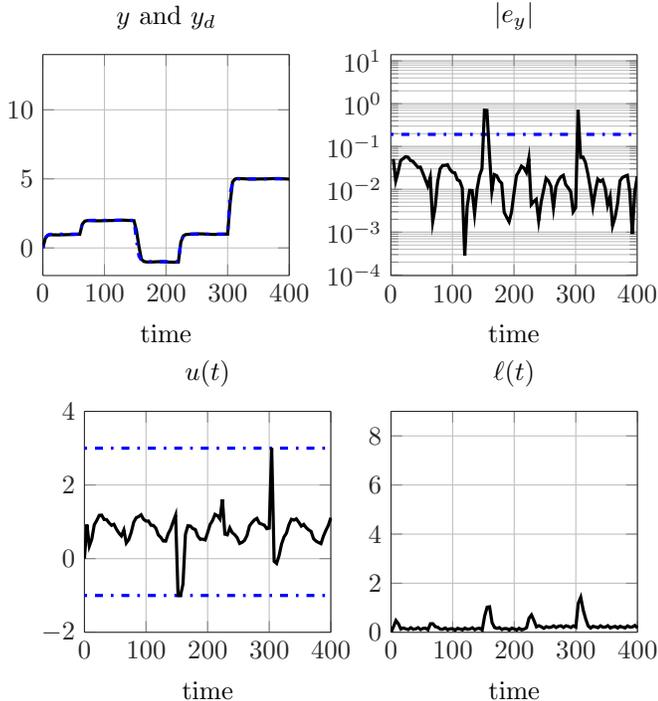

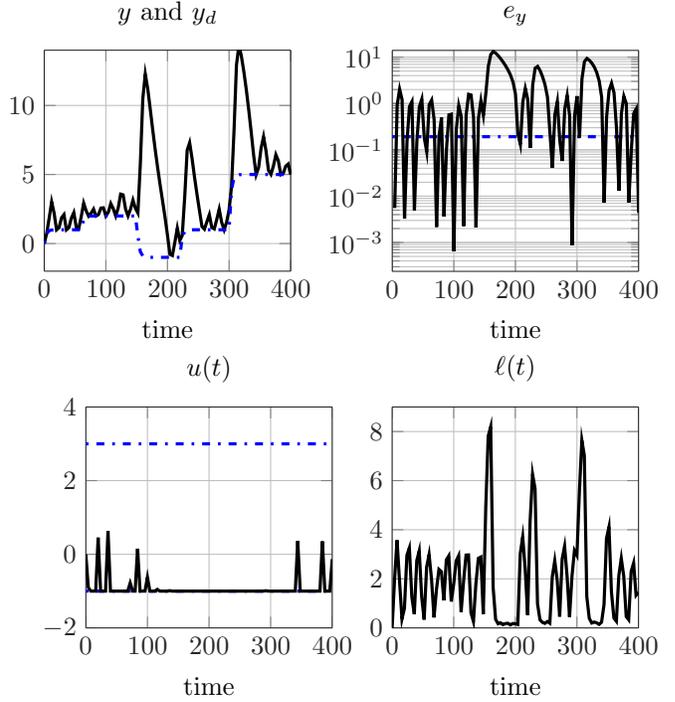
\begin{figure}
\begin{center}
\input{figsimbad.tex}
\end{center} 
\caption{Behavior of the closed-loop system for the choice $(\lambda^{(2)},\lambda_f)=(170,0.0102)$ that violate the prescribed bounds.} \label{figsimbad} 
\end{figure}

Let us consider the uncertain system given by (\ref{syst1})-(\ref{syst2}) in which the internal dynamic (\ref{syst2}) is given by:
\begin{align*}
\dot\eta(t)&=A\eta(t)+B_1\|\dot{\bm u}^{[t-\tau,t]}\|_\infty +B_2w\\
\ell(\eta)&=C\eta
\end{align*} 
where $A$, $B_1$, $B_2$ and $C$ are given by:
\begin{align*}
A&= \begin{bmatrix}
-3.4& -1 &-0.5\cr 0.25 &-1.7 &0.5\cr 1.2& 2.75& -3.6
\end{bmatrix}\ ;\ B_1= \begin{bmatrix}
0.1\cr 0\cr 0
\end{bmatrix}\ ;\ B_2= \begin{bmatrix}
0\cr 0.2\cr 0
\end{bmatrix}    \\
C&= \begin{bmatrix}
1&0&1
\end{bmatrix}  
\end{align*}
while the delay $\tau=10$ is used. The unknown signal $g(t)$ is given by:
\begin{equation}
g(t):=1-0.3\cos(0.1t)-0.1\sin(0.2t+\pi/6);
\end{equation} 
Therefore, the bounds $\underline g=0.5$ and $\bar g=1.5$ can be used ($\Delta_g=1$). Moreover, the bound on the derivative of $g$ is $\delta_g=0.05$. Therefore, considering control saturations given by 
$\underline u=-1$ and $\bar u=3$ leads to $\varrho^-=1.5$ and $\varrho^+=1.5$, $\underline\varrho=1.5$ and $\Delta_u=4$. The bounds $\underline \alpha=0.1$, $\bar \alpha=0.3$ are supposed to hold while a {\em true} value $\alpha=0.2$ is used in the simulations. We consider as admissible set $\mathbb W$ to be the set of all disturbances that are bounded by $\bar w=0.1$. For simulation, the disturbance signal $w(t)=\bar w\sin(t)$ is used. Computation of (\ref{defdec0lin})-(\ref{defdec1lin}) gives:
\begin{equation}
c_0 =0.05 \;\ c_1 =0.07\ ;\ d_0 =0.2\ ;\ d_1 =0.42
\end{equation} 
Using these values in the definition of $\varphi(\cdot)$ and $\lambda^*$ leads to the evolution of $\varphi(\lambda)$, the bound on the asymptotic tracking error and the bound $\lambda\varphi(\lambda)$ on the values of $\lambda_f$ are depicted in Figure \ref{analysis_plots} where it can be inferred that $\lambda^*=17.18$. Note that the upper bound $1/(\bar\alpha c_1)\approx 47.4$ in this example meaning that in this example, $\lambda^*$ is limited by the need for a positive sign of $\varphi$.\ \\ \ \\ 
Figure \ref{figsimgood} shows the behavior of the closed-loop system when the value $\lambda^{(1)}=17\in (0,\lambda^*)$ is adopted together with $\lambda_f=0.95\lambda\varphi(\lambda)\approx 0.0102$ satisfying (\ref{condonlambdaf}). This Figure shows quite good closed-loop behavior with the tracking error asymptotically below its theoretical bound a while after the set-point gets constant. \ \\ \ \\ 
On the other hand, Figure \ref{figsimbad}  shows the behavior of the closed-loop system when $\lambda^{(2)}=10\lambda^{(1)}$ is used (with the same value of $\lambda_f$. The closed-loop clearly shows that the trajectories are no more adequately attracted to a small neighborhood of the desired values. This is mainly due to the high excursion of the internal state $\eta$ which leads to high value of $\ell$ that breaks the attractiveness of the tight neighborhood of $y_d$. This excursion has to be compared to the rather moderate one which can be observed on Figure \ref{figsimgood}. This clearly shows that a rationalized choice of $(\lambda,\lambda_f)$ is a key step in the success of the simple control law (\ref{defdeu})-(\ref{defdezdot}) for this particular class of systems. 
\section{Conclusion and future works} \label{secconc} 
In this paper, the problem of dynamic output feedback of systems with control-derivative dependent uncertainties is analyzed. High level characterization of the involved map is given so that a successful tuning of a simple feedback law is given together with an upper bound on the output tracking error.  A natural extension of this work is a concrete application to the realistic examples described in the paper. 

\bibliographystyle{plain}
\bibliography{biblio_MIMO_uncertain.bib}    
\end{document}

%% file: figsimgood.tex
%
%
\begin{tikzpicture}

\begin{axis}[%
width=0.18\textwidth,
height=0.12\textheight,
scale only axis,
separate axis lines,
every outer x axis line/.append style={darkgray!60!black},
every x tick label/.append style={font=\color{darkgray!60!black}},
xmin=0,
xmax=400,
xlabel={time},
xmajorgrids,
every outer y axis line/.append style={darkgray!60!black},
every y tick label/.append style={font=\color{darkgray!60!black}},
ymin=-2,
ymax=14,
ymajorgrids,
name=plot1,
title={$y$ and $y_d$}
]
\addplot [
color=black,
solid,
line width=1.2,
]
table[row sep=crcr]{
0 0\\
4 0.685324201821661\\
8 0.914803002194244\\
12 0.955717710752813\\
16 0.947733930679313\\
20 0.946597523615868\\
24 0.941562279286519\\
28 0.943682363768402\\
32 0.95300524829434\\
36 0.954949240717198\\
40 0.961764275067334\\
44 0.967313441125401\\
48 0.966921075851842\\
52 0.97595957776374\\
56 0.983055211699095\\
60 0.98800672717671\\
64 1.71715741021058\\
68 1.93226747693184\\
72 1.97782963354112\\
76 1.98022484502818\\
80 1.9686854622135\\
84 1.9636386782521\\
88 1.96365038496565\\
92 1.96231465781557\\
96 1.9707353994208\\
100 1.97490877236174\\
104 1.97571694912641\\
108 1.98328059384125\\
112 1.98432799369279\\
116 1.98925113034975\\
120 2.00028467994207\\
124 2.00403205403202\\
128 2.00946009315543\\
132 2.00935280531326\\
136 1.99729490981421\\
140 1.98996262620276\\
144 1.97924313181845\\
148 1.96972177053926\\
152 1.2433978266999\\
156 0.103131080409842\\
160 -0.822080466822602\\
164 -0.980943767391157\\
168 -1.01466618936718\\
172 -1.0188008185135\\
176 -1.01536599156778\\
180 -1.01376010243907\\
184 -1.00288440134989\\
188 -0.997749806900158\\
192 -0.998239133683847\\
196 -0.997023994475911\\
200 -1.00834577907869\\
204 -1.02099954539364\\
208 -1.02630765910587\\
212 -1.03492317766818\\
216 -1.03154174920412\\
220 -1.02491150867568\\
224 0.421184935800992\\
228 0.865450543027546\\
232 0.968569512723855\\
236 0.984512534472595\\
240 0.999320282921591\\
244 1.00435785111881\\
248 1.00934186441949\\
252 1.01805498369684\\
256 1.01567066608755\\
260 1.01173716694178\\
264 1.00372759961685\\
268 0.988571989879382\\
272 0.983890885781331\\
276 0.980394965705361\\
280 0.978727975714028\\
284 0.987188285683778\\
288 0.989163240599932\\
292 0.991146883806566\\
296 0.997004482441994\\
300 0.996293108927289\\
304 3.23369398291075\\
308 4.77812623968019\\
312 4.98343693144507\\
316 5.02283283009798\\
320 5.01752991932434\\
324 5.00920411626794\\
328 5.00369636006443\\
332 4.99059540056788\\
336 4.98260412265063\\
340 4.98469629076181\\
344 4.98331204645862\\
348 4.98926757687851\\
352 4.99577546683138\\
356 4.99325688722104\\
360 4.99858320057734\\
364 5.00221074298923\\
368 5.00454104464625\\
372 5.01580513107447\\
376 5.0188981716727\\
380 5.02027209526032\\
384 5.02125413630614\\
388 5.00861655352423\\
392 4.99908942379822\\
396 4.99140791722968\\
400 4.9798400046612\\
};
\addplot [
color=blue,
dash pattern=on 1pt off 3pt on 3pt off 3pt,
line width=1.2,
]
table[row sep=crcr]{
0 0\\
4 0.736402861884272\\
8 0.930516548777199\\
12 0.981684361111266\\
16 0.995172050006168\\
20 0.998727366198658\\
24 0.999664537372096\\
28 0.99991157301134\\
32 0.999976690898858\\
36 0.999993855787645\\
40 0.999998380403207\\
44 0.99999957307892\\
48 0.999999887464826\\
52 0.99999997033605\\
56 0.999999992180668\\
60 0.999999997938847\\
64 1.73640286134096\\
68 1.93051654863398\\
72 1.98168436107352\\
76 1.99517204999622\\
80 1.99872736619604\\
84 1.9996645373714\\
88 1.99991157301116\\
92 1.99997669089881\\
96 1.99999385578764\\
100 1.9999983804032\\
104 1.99999957307892\\
108 1.99999988746483\\
112 1.99999997033605\\
116 1.99999999218066\\
120 1.99999999793885\\
124 1.99999999945669\\
128 1.99999999985679\\
132 1.99999999996225\\
136 1.99999999999005\\
140 1.99999999999738\\
144 1.99999999999931\\
148 1.99999999999982\\
152 0.540251357097618\\
156 -0.593994150290206\\
160 -0.892978019958268\\
164 -0.97178931234552\\
168 -0.992563743470002\\
172 -0.998039824060398\\
176 -0.999483303232117\\
180 -0.999863800210712\\
184 -0.999964098125331\\
188 -0.999990536368584\\
192 -0.999997505413843\\
196 -0.999999342434227\\
200 -0.999999826667546\\
204 -0.999999954310062\\
208 -0.999999987956263\\
212 -0.999999996825306\\
216 -0.999999999163159\\
220 -0.999999999779392\\
224 0.47280572382672\\
228 0.861033097569721\\
232 0.963368722226571\\
236 0.990344100013402\\
240 0.9974547323976\\
244 0.99932907474427\\
248 0.999823146022699\\
252 0.99995338179772\\
256 0.999987711575296\\
260 0.999996760806416\\
264 0.999999146157843\\
268 0.999999774929652\\
272 0.999999940672101\\
276 0.999999984361336\\
280 0.999999995877692\\
284 0.999999998913371\\
288 0.999999999713568\\
292 0.999999999924499\\
296 0.999999999980099\\
300 0.999999999994755\\
304 3.9456114475358\\
308 4.7220661951085\\
312 4.92673744444497\\
316 4.98068820002465\\
320 4.99490946479463\\
324 4.99865814948839\\
328 4.99964629204535\\
332 4.99990676359543\\
336 4.99997542315059\\
340 4.99999352161283\\
344 4.99999829231568\\
348 4.9999995498593\\
352 4.9999998813442\\
356 4.99999996872268\\
360 4.99999999175539\\
364 4.99999999782674\\
368 4.99999999942713\\
372 4.999999999849\\
376 4.9999999999602\\
380 4.99999999998951\\
384 4.99999999999723\\
388 4.99999999999927\\
392 4.99999999999978\\
396 4.9999999999998\\
400 4.9999999999998\\
};
\end{axis}

\begin{axis}[%
width=0.18\textwidth,
height=0.12\textheight,
scale only axis,
separate axis lines,
every outer x axis line/.append style={darkgray!60!black},
every x tick label/.append style={font=\color{darkgray!60!black}},
xmin=0,
xmax=400,
xlabel={time},
xmajorgrids,
every outer y axis line/.append style={darkgray!60!black},
every y tick label/.append style={font=\color{darkgray!60!black}},
ymode=log,
ymin=0,
ymax=14,
yminorticks=true,
ytick={0.0001,0.001,0.01,0.1,1,10},
ymajorgrids,
yminorgrids,
name=plot2,
at=(plot1.right of south east),
anchor=left of south west,
title={$\vert e_y\vert$}
]
\addplot [
color=blue,
dash pattern=on 1pt off 3pt on 3pt off 3pt,
line width=1.2,
]
table[row sep=crcr]{
0 0.192\\
4 0.192\\
8 0.192\\
12 0.192\\
16 0.192\\
20 0.192\\
24 0.192\\
28 0.192\\
32 0.192\\
36 0.192\\
40 0.192\\
44 0.192\\
48 0.192\\
52 0.192\\
56 0.192\\
60 0.192\\
64 0.192\\
68 0.192\\
72 0.192\\
76 0.192\\
80 0.192\\
84 0.192\\
88 0.192\\
92 0.192\\
96 0.192\\
100 0.192\\
104 0.192\\
108 0.192\\
112 0.192\\
116 0.192\\
120 0.192\\
124 0.192\\
128 0.192\\
132 0.192\\
136 0.192\\
140 0.192\\
144 0.192\\
148 0.192\\
152 0.192\\
156 0.192\\
160 0.192\\
164 0.192\\
168 0.192\\
172 0.192\\
176 0.192\\
180 0.192\\
184 0.192\\
188 0.192\\
192 0.192\\
196 0.192\\
200 0.192\\
204 0.192\\
208 0.192\\
212 0.192\\
216 0.192\\
220 0.192\\
224 0.192\\
228 0.192\\
232 0.192\\
236 0.192\\
240 0.192\\
244 0.192\\
248 0.192\\
252 0.192\\
256 0.192\\
260 0.192\\
264 0.192\\
268 0.192\\
272 0.192\\
276 0.192\\
280 0.192\\
284 0.192\\
288 0.192\\
292 0.192\\
296 0.192\\
300 0.192\\
304 0.192\\
308 0.192\\
312 0.192\\
316 0.192\\
320 0.192\\
324 0.192\\
328 0.192\\
332 0.192\\
336 0.192\\
340 0.192\\
344 0.192\\
348 0.192\\
352 0.192\\
356 0.192\\
360 0.192\\
364 0.192\\
368 0.192\\
372 0.192\\
376 0.192\\
380 0.192\\
384 0.192\\
388 0.192\\
392 0.192\\
396 0.192\\
400 0.192\\
};
\addplot [
color=black,
solid,
line width=1.2,
]
table[row sep=crcr]{
0 0\\
4 0.051078660062611\\
8 0.0157135465829552\\
12 0.0259666503584526\\
16 0.0474381193268548\\
20 0.0521298425827896\\
24 0.0581022580855772\\
28 0.0562292092429374\\
32 0.0469714426045182\\
36 0.0450446150704474\\
40 0.0382341053358731\\
44 0.0326861319535194\\
48 0.0330788116129848\\
52 0.0240403925723098\\
56 0.0169447804815733\\
60 0.0119932707621365\\
64 0.0192454511303795\\
68 0.0017509282978585\\
72 0.0038547275323999\\
76 0.0149472049680408\\
80 0.0300419039825399\\
84 0.0360258591193021\\
88 0.0362611880455048\\
92 0.0376620330832378\\
96 0.0292584563668419\\
100 0.0250896080414658\\
104 0.024282623952504\\
108 0.0167192936235776\\
112 0.0156719766432547\\
116 0.0107488618309159\\
120 0.000284682003218073\\
124 0.00403205457533273\\
128 0.0094600932986455\\
132 0.00935280535101057\\
136 0.0027050901758412\\
140 0.0100373737946156\\
144 0.020756868180857\\
148 0.030278229460561\\
152 0.703146469602278\\
156 0.697125230700048\\
160 0.0708975531356655\\
164 0.00915445504563739\\
168 0.0221024458971752\\
172 0.0207609944531042\\
176 0.0158826883356626\\
180 0.0138963022283557\\
184 0.00292030322455805\\
188 0.00224072946842624\\
192 0.00175837172999571\\
196 0.00297534795831611\\
200 0.00834595241114466\\
204 0.0209995910835798\\
208 0.0263076711496062\\
212 0.0349231808428778\\
216 0.0315417500409568\\
220 0.0249115088962925\\
224 0.0516207880257281\\
228 0.00441744545782485\\
232 0.00520079049728384\\
236 0.00583156554080733\\
240 0.00186555052399118\\
244 0.00502877637454513\\
248 0.00951871839678797\\
252 0.01810160189912\\
256 0.0156829545122563\\
260 0.0117404061353689\\
264 0.00372845345900485\\
268 0.0114277850502692\\
272 0.0161090548907701\\
276 0.0196050186559749\\
280 0.0212720201636644\\
284 0.0128117132295938\\
288 0.0108367591136352\\
292 0.00885311611793316\\
296 0.00299551753810445\\
300 0.00370689106746647\\
304 0.71191746462505\\
308 0.0560600445716899\\
312 0.0566994870001043\\
316 0.042144630073337\\
320 0.0226204545297035\\
324 0.0105459667795511\\
328 0.00405006801908048\\
332 0.00931136302754432\\
336 0.0173713004999545\\
340 0.0152972308510195\\
344 0.0166862458570565\\
348 0.010731972980798\\
352 0.00422441451281674\\
356 0.00674308150164027\\
360 0.00141679117804649\\
364 0.00221074516249153\\
368 0.00454104521912235\\
372 0.0158051312254734\\
376 0.018898171712503\\
380 0.0202720952708138\\
384 0.0212541363089107\\
388 0.00861655352496005\\
392 0.000910576201561319\\
396 0.00859208277011358\\
400 0.0201599953386014\\
};
\end{axis}

\begin{axis}[%
width=0.18\textwidth,
height=0.12\textheight,
scale only axis,
separate axis lines,
every outer x axis line/.append style={darkgray!60!black},
every x tick label/.append style={font=\color{darkgray!60!black}},
xmin=0,
xmax=400,
xlabel={time},
xmajorgrids,
every outer y axis line/.append style={darkgray!60!black},
every y tick label/.append style={font=\color{darkgray!60!black}},
ymin=0,
ymax=9,
ymajorgrids,
name=plot4,
at=(plot2.below south west),
anchor=above north west,
title={$\ell(t)$}
]
\addplot [
color=black,
solid,
line width=1.2,
]
table[row sep=crcr]{
0 0\\
4 0.228143461123392\\
8 0.479043955821622\\
12 0.347365401510214\\
16 0.130756254640908\\
20 0.177650890837441\\
24 0.139694875000125\\
28 0.111791232790287\\
32 0.187597464035459\\
36 0.123817312571612\\
40 0.138556133100422\\
44 0.190077631112405\\
48 0.112145619955744\\
52 0.158962722240794\\
56 0.170128251596067\\
60 0.102126995165625\\
64 0.348785266580453\\
68 0.350324115532313\\
72 0.226974634469689\\
76 0.20540478911556\\
80 0.126858854388477\\
84 0.125562691669524\\
88 0.176685156737195\\
92 0.101781631329478\\
96 0.143378991648268\\
100 0.149414131741843\\
104 0.0838861902649659\\
108 0.162721208432971\\
112 0.134700623753941\\
116 0.100855908699925\\
120 0.172469980933542\\
124 0.123146134318489\\
128 0.134383223932465\\
132 0.185414268820903\\
136 0.10622020999756\\
140 0.153354903988268\\
144 0.164503466441384\\
148 0.0919306019666643\\
152 0.664622774707217\\
156 1.0221216437442\\
160 1.0406822850664\\
164 0.399841117458701\\
168 0.195916103845448\\
172 0.159314737311708\\
176 0.193157728618849\\
180 0.112347609360571\\
184 0.154023510614563\\
188 0.149770803224534\\
192 0.0901915852982101\\
196 0.169300617936118\\
200 0.140400692771185\\
204 0.090599494030858\\
208 0.164577119647352\\
212 0.101002080185803\\
216 0.128309070442871\\
220 0.186688667336642\\
224 0.577059026975683\\
228 0.709139834396679\\
232 0.482478448299855\\
236 0.223194786187991\\
240 0.270635533117228\\
244 0.225091057323535\\
248 0.161290306311469\\
252 0.232563150073484\\
256 0.196692794376433\\
260 0.208013865461937\\
264 0.264315787931586\\
268 0.19512529697622\\
272 0.2457578144704\\
276 0.265529433576083\\
280 0.197673083960631\\
284 0.265276926312168\\
288 0.238548965340764\\
292 0.19566889993346\\
296 0.26376642380422\\
300 0.212385071098202\\
304 1.18926391717328\\
308 1.42629926220318\\
312 0.903245510006981\\
316 0.545019962723524\\
320 0.259629000181423\\
324 0.1719798817359\\
328 0.245443302027847\\
332 0.220186357851887\\
336 0.187491085786282\\
340 0.265173393531616\\
344 0.208406229095502\\
348 0.212351370790865\\
352 0.269783160000785\\
356 0.1753060189458\\
360 0.215658521942422\\
364 0.242312173237542\\
368 0.175200598257583\\
372 0.244119007449092\\
376 0.208552510532549\\
380 0.17721863120282\\
384 0.261668749406119\\
388 0.208609429868173\\
392 0.220782067039308\\
396 0.280553235444786\\
400 0.17840866325832\\
};
\end{axis}

\begin{axis}[%
width=0.18\textwidth,
height=0.12\textheight,
scale only axis,
separate axis lines,
every outer x axis line/.append style={darkgray!60!black},
every x tick label/.append style={font=\color{darkgray!60!black}},
xmin=0,
xmax=400,
xlabel={time},
xmajorgrids,
every outer y axis line/.append style={darkgray!60!black},
every y tick label/.append style={font=\color{darkgray!60!black}},
ymin=-2,
ymax=4,
ymajorgrids,
at=(plot4.left of south west),
anchor=right of south east,
title={$u(t)$}
]
\addplot [
color=blue,
dash pattern=on 1pt off 3pt on 3pt off 3pt,
line width=1.2,
]
table[row sep=crcr]{
0 -1\\
4 -1\\
8 -1\\
12 -1\\
16 -1\\
20 -1\\
24 -1\\
28 -1\\
32 -1\\
36 -1\\
40 -1\\
44 -1\\
48 -1\\
52 -1\\
56 -1\\
60 -1\\
64 -1\\
68 -1\\
72 -1\\
76 -1\\
80 -1\\
84 -1\\
88 -1\\
92 -1\\
96 -1\\
100 -1\\
104 -1\\
108 -1\\
112 -1\\
116 -1\\
120 -1\\
124 -1\\
128 -1\\
132 -1\\
136 -1\\
140 -1\\
144 -1\\
148 -1\\
152 -1\\
156 -1\\
160 -1\\
164 -1\\
168 -1\\
172 -1\\
176 -1\\
180 -1\\
184 -1\\
188 -1\\
192 -1\\
196 -1\\
200 -1\\
204 -1\\
208 -1\\
212 -1\\
216 -1\\
220 -1\\
224 -1\\
228 -1\\
232 -1\\
236 -1\\
240 -1\\
244 -1\\
248 -1\\
252 -1\\
256 -1\\
260 -1\\
264 -1\\
268 -1\\
272 -1\\
276 -1\\
280 -1\\
284 -1\\
288 -1\\
292 -1\\
296 -1\\
300 -1\\
304 -1\\
308 -1\\
312 -1\\
316 -1\\
320 -1\\
324 -1\\
328 -1\\
332 -1\\
336 -1\\
340 -1\\
344 -1\\
348 -1\\
352 -1\\
356 -1\\
360 -1\\
364 -1\\
368 -1\\
372 -1\\
376 -1\\
380 -1\\
384 -1\\
388 -1\\
392 -1\\
396 -1\\
400 -1\\
};
\addplot [
color=blue,
dash pattern=on 1pt off 3pt on 3pt off 3pt,
line width=1.2,
]
table[row sep=crcr]{
0 3\\
4 3\\
8 3\\
12 3\\
16 3\\
20 3\\
24 3\\
28 3\\
32 3\\
36 3\\
40 3\\
44 3\\
48 3\\
52 3\\
56 3\\
60 3\\
64 3\\
68 3\\
72 3\\
76 3\\
80 3\\
84 3\\
88 3\\
92 3\\
96 3\\
100 3\\
104 3\\
108 3\\
112 3\\
116 3\\
120 3\\
124 3\\
128 3\\
132 3\\
136 3\\
140 3\\
144 3\\
148 3\\
152 3\\
156 3\\
160 3\\
164 3\\
168 3\\
172 3\\
176 3\\
180 3\\
184 3\\
188 3\\
192 3\\
196 3\\
200 3\\
204 3\\
208 3\\
212 3\\
216 3\\
220 3\\
224 3\\
228 3\\
232 3\\
236 3\\
240 3\\
244 3\\
248 3\\
252 3\\
256 3\\
260 3\\
264 3\\
268 3\\
272 3\\
276 3\\
280 3\\
284 3\\
288 3\\
292 3\\
296 3\\
300 3\\
304 3\\
308 3\\
312 3\\
316 3\\
320 3\\
324 3\\
328 3\\
332 3\\
336 3\\
340 3\\
344 3\\
348 3\\
352 3\\
356 3\\
360 3\\
364 3\\
368 3\\
372 3\\
376 3\\
380 3\\
384 3\\
388 3\\
392 3\\
396 3\\
400 3\\
};
\addplot [
color=black,
solid,
line width=1.2,
]
table[row sep=crcr]{
0 0\\
4 0.925062497553322\\
8 0.342758842380343\\
12 0.530236230354081\\
16 0.920312442475334\\
20 1.03483177299581\\
24 1.17595473849203\\
28 1.18241862324173\\
32 1.06155323400026\\
36 1.06096125183122\\
40 0.972566606722455\\
44 0.903995286831982\\
48 0.933049377215702\\
52 0.798480836315891\\
56 0.693524359272014\\
60 0.618205623452547\\
64 0.777843301632089\\
68 0.425832526791268\\
72 0.518686973386126\\
76 0.714844939713628\\
80 0.987582249936211\\
84 1.11103335379545\\
88 1.14170799177169\\
92 1.19090689142695\\
96 1.070326220209\\
100 1.01967085241535\\
104 1.02188312877904\\
108 0.907402883687396\\
112 0.901988039888438\\
116 0.826144593857118\\
120 0.642960536971306\\
124 0.57812526668605\\
128 0.479556244274843\\
132 0.475777579182645\\
136 0.677874545315103\\
140 0.806015851038203\\
144 1.00044725597614\\
148 1.17915047259061\\
152 -1\\
156 -1\\
160 -0.690383466283512\\
164 0.661983441649853\\
168 0.894563351157517\\
172 0.885206840415235\\
176 0.816459266447698\\
180 0.792792678602474\\
184 0.611192339199622\\
188 0.524838398380554\\
192 0.530540480793797\\
196 0.507711373815486\\
200 0.70292067759791\\
204 0.926899704024873\\
208 1.03436230374157\\
212 1.20277183063225\\
216 1.16715485942944\\
220 1.07528807394706\\
224 1.60694828215275\\
228 0.665656814392278\\
232 0.646760777160285\\
236 0.834932215562921\\
240 0.705571705884806\\
244 0.650498921352249\\
248 0.567887654872883\\
252 0.412909283917859\\
256 0.442527246966769\\
260 0.498344589258405\\
264 0.630263624806462\\
268 0.890451252284036\\
272 0.978853848328496\\
276 1.0523570931788\\
280 1.09407819281955\\
284 0.961922316413973\\
288 0.937664605038258\\
292 0.90928374546107\\
296 0.814341819923487\\
300 0.829501369004751\\
304 3\\
308 -0.0735152807898781\\
312 -0.130497359161695\\
316 0.0837157601006414\\
320 0.395567203953889\\
324 0.58860090275407\\
328 0.693672307180474\\
332 0.923916628224781\\
336 1.06910855739163\\
340 1.04602964750037\\
344 1.0814877625705\\
348 0.988388209809527\\
352 0.884061834560416\\
356 0.930345376244674\\
360 0.841761030923642\\
364 0.781372392376809\\
368 0.738601214499547\\
372 0.539899370525442\\
376 0.476382963126187\\
380 0.438103959148863\\
384 0.407237108781501\\
388 0.61214382882554\\
392 0.769960993852052\\
396 0.905191538711948\\
400 1.11177622513009\\
};
\end{axis}
\end{tikzpicture}%

%% file: figsimbad.tex
%
%
\begin{tikzpicture}

\begin{axis}[%
width=0.18\textwidth,
height=0.12\textheight,
scale only axis,
separate axis lines,
every outer x axis line/.append style={darkgray!60!black},
every x tick label/.append style={font=\color{darkgray!60!black}},
xmin=0,
xmax=400,
xlabel={time},
xmajorgrids,
every outer y axis line/.append style={darkgray!60!black},
every y tick label/.append style={font=\color{darkgray!60!black}},
ymin=-2,
ymax=14,
ymajorgrids,
name=plot1,
title={$y$ and $y_d$}
]
\addplot [
color=black,
solid,
line width=1.2,
]
table[row sep=crcr]{
0 0\\
4 0.741948597036438\\
8 1.8517619923127\\
12 3.01573229114769\\
16 2.208367841231\\
20 0.995458648313025\\
24 1.15148800276509\\
28 1.86065400869124\\
32 2.09641906531023\\
36 0.995000555539633\\
40 1.15219523394979\\
44 1.84563574146639\\
48 2.24059511404762\\
52 1.18081738634705\\
56 1.09072136325528\\
60 1.94162504405823\\
64 2.92357417340598\\
68 2.37816799883446\\
72 1.98385188157074\\
76 2.19664760304021\\
80 2.49565879845895\\
84 1.99602865571819\\
88 2.09884378949546\\
92 2.56867311315762\\
96 2.58843524714794\\
100 2.00064589341889\\
104 2.27054500283322\\
108 2.89608831382556\\
112 2.52502686154344\\
116 2.00226321980519\\
120 2.56825436477775\\
124 3.578627719258\\
128 3.52623235371631\\
132 2.49968435118472\\
136 2.00210551218541\\
140 2.48962436706072\\
144 3.14338856233816\\
148 2.55076278059988\\
152 2.2144466311279\\
156 5.80067588410222\\
160 10.4921419871392\\
164 12.2687450495288\\
168 11.2032837513559\\
172 9.76010602560803\\
176 8.26359109022355\\
180 6.84001316089595\\
184 5.52386124416225\\
188 4.27610597035375\\
192 3.10857647950916\\
196 1.92074377588219\\
200 0.622718552683112\\
204 -0.781319400639395\\
208 -0.869044204826687\\
212 0.0895480617876436\\
216 1.11554395466719\\
220 0.219799265979029\\
224 0.58246436429777\\
228 3.35952384117719\\
232 6.75620638558209\\
236 7.30257508829626\\
240 6.14318960719819\\
244 4.78994172242985\\
248 3.54595023950421\\
252 2.37566479577091\\
256 1.23699761657084\\
260 1.0408789133818\\
264 1.57338957170571\\
268 2.0901601026556\\
272 1.15881946892394\\
276 1.05743802382321\\
280 1.74912272436729\\
284 2.48633863283533\\
288 1.53493577362824\\
292 1.00086612388171\\
296 1.54492633360168\\
300 2.44041773656682\\
304 4.12588808818683\\
308 8.12748949974977\\
312 12.9131350068114\\
316 14.346685268445\\
320 13.571864814406\\
324 12.4435027034603\\
328 11.1383054798913\\
332 9.64584614132187\\
336 8.04580822515519\\
340 6.35420628498455\\
344 4.99276563235249\\
348 5.51446055607671\\
352 6.90953454168216\\
356 7.73735559228558\\
360 6.70850968395953\\
364 5.30045206107737\\
368 5.0130749519301\\
372 5.61119778550408\\
376 6.43188707642497\\
380 5.99221727280001\\
384 4.99271847875113\\
388 5.09279711311583\\
392 5.6076228730044\\
396 5.76141185008172\\
400 4.99563215072452\\
};
\addplot [
color=blue,
dash pattern=on 1pt off 3pt on 3pt off 3pt,
line width=1.2,
]
table[row sep=crcr]{
0 0\\
4 0.736402861884274\\
8 0.930516548777198\\
12 0.981684361111265\\
16 0.995172050006167\\
20 0.998727366198661\\
24 0.999664537372098\\
28 0.99991157301134\\
32 0.999976690898858\\
36 0.999993855787648\\
40 0.999998380403207\\
44 0.999999573078921\\
48 0.999999887464826\\
52 0.99999997033605\\
56 0.999999992180666\\
60 0.999999997938847\\
64 1.73640286134097\\
68 1.93051654863398\\
72 1.98168436107352\\
76 1.99517204999622\\
80 1.99872736619604\\
84 1.99966453737141\\
88 1.99991157301116\\
92 1.99997669089881\\
96 1.99999385578763\\
100 1.9999983804032\\
104 1.99999957307892\\
108 1.99999988746483\\
112 1.99999997033605\\
116 1.99999999218067\\
120 1.99999999793884\\
124 1.99999999945668\\
128 1.99999999985678\\
132 1.99999999996225\\
136 1.99999999999005\\
140 1.99999999999738\\
144 1.99999999999931\\
148 1.99999999999982\\
152 0.540251357097689\\
156 -0.593994150290161\\
160 -0.892978019958235\\
164 -0.971789312345509\\
168 -0.992563743469999\\
172 -0.998039824060396\\
176 -0.999483303232117\\
180 -0.999863800210713\\
184 -0.999964098125333\\
188 -0.999990536368585\\
192 -0.999997505413843\\
196 -0.999999342434229\\
200 -0.999999826667544\\
204 -0.999999954310061\\
208 -0.999999987956262\\
212 -0.999999996825306\\
216 -0.999999999163159\\
220 -0.999999999779392\\
224 0.472805723826688\\
228 0.861033097569708\\
232 0.963368722226558\\
236 0.990344100013398\\
240 0.997454732397599\\
244 0.999329074744269\\
248 0.999823146022699\\
252 0.999953381797719\\
256 0.999987711575296\\
260 0.999996760806415\\
264 0.999999146157843\\
268 0.99999977492965\\
272 0.999999940672099\\
276 0.999999984361335\\
280 0.999999995877693\\
284 0.999999998913372\\
288 0.999999999713567\\
292 0.999999999924497\\
296 0.999999999980098\\
300 0.999999999994753\\
304 3.94561144753587\\
308 4.72206619510849\\
312 4.92673744444499\\
316 4.98068820002466\\
320 4.99490946479463\\
324 4.99865814948838\\
328 4.99964629204535\\
332 4.99990676359542\\
336 4.99997542315059\\
340 4.99999352161283\\
344 4.99999829231569\\
348 4.9999995498593\\
352 4.9999998813442\\
356 4.99999996872267\\
360 4.99999999175538\\
364 4.99999999782675\\
368 4.99999999942713\\
372 4.99999999984899\\
376 4.99999999996019\\
380 4.99999999998951\\
384 4.99999999999723\\
388 4.99999999999927\\
392 4.99999999999977\\
396 4.9999999999998\\
400 4.99999999999981\\
};
\end{axis}

\begin{axis}[%
width=0.18\textwidth,
height=0.12\textheight,
scale only axis,
separate axis lines,
every outer x axis line/.append style={darkgray!60!black},
every x tick label/.append style={font=\color{darkgray!60!black}},
xmin=0,
xmax=400,
xlabel={time},
xmajorgrids,
every outer y axis line/.append style={darkgray!60!black},
every y tick label/.append style={font=\color{darkgray!60!black}},
ymode=log,
ymin=0,
ymax=14,
ytick={0.0001,0.001,0.01,0.1,1,10},
yminorticks=true,
ymajorgrids,
yminorgrids,
name=plot2,
at=(plot1.right of south east),
anchor=left of south west,
title={$e_y$}
]
\addplot [
color=blue,
dash pattern=on 1pt off 3pt on 3pt off 3pt,
line width = 1.2,
]
table[row sep=crcr]{
0 0.192\\
4 0.192\\
8 0.192\\
12 0.192\\
16 0.192\\
20 0.192\\
24 0.192\\
28 0.192\\
32 0.192\\
36 0.192\\
40 0.192\\
44 0.192\\
48 0.192\\
52 0.192\\
56 0.192\\
60 0.192\\
64 0.192\\
68 0.192\\
72 0.192\\
76 0.192\\
80 0.192\\
84 0.192\\
88 0.192\\
92 0.192\\
96 0.192\\
100 0.192\\
104 0.192\\
108 0.192\\
112 0.192\\
116 0.192\\
120 0.192\\
124 0.192\\
128 0.192\\
132 0.192\\
136 0.192\\
140 0.192\\
144 0.192\\
148 0.192\\
152 0.192\\
156 0.192\\
160 0.192\\
164 0.192\\
168 0.192\\
172 0.192\\
176 0.192\\
180 0.192\\
184 0.192\\
188 0.192\\
192 0.192\\
196 0.192\\
200 0.192\\
204 0.192\\
208 0.192\\
212 0.192\\
216 0.192\\
220 0.192\\
224 0.192\\
228 0.192\\
232 0.192\\
236 0.192\\
240 0.192\\
244 0.192\\
248 0.192\\
252 0.192\\
256 0.192\\
260 0.192\\
264 0.192\\
268 0.192\\
272 0.192\\
276 0.192\\
280 0.192\\
284 0.192\\
288 0.192\\
292 0.192\\
296 0.192\\
300 0.192\\
304 0.192\\
308 0.192\\
312 0.192\\
316 0.192\\
320 0.192\\
324 0.192\\
328 0.192\\
332 0.192\\
336 0.192\\
340 0.192\\
344 0.192\\
348 0.192\\
352 0.192\\
356 0.192\\
360 0.192\\
364 0.192\\
368 0.192\\
372 0.192\\
376 0.192\\
380 0.192\\
384 0.192\\
388 0.192\\
392 0.192\\
396 0.192\\
400 0.192\\
};
\addplot [
color=black,
solid,
line width=1.2,
]
table[row sep=crcr]{
0 0\\
4 0.00554573515216383\\
8 0.921245443535504\\
12 2.03404793003643\\
16 1.21319579122483\\
20 0.0032687178856361\\
24 0.151823465392992\\
28 0.860742435679902\\
32 1.09644237441137\\
36 0.00499330024801525\\
40 0.152196853546581\\
44 0.845636168387468\\
48 1.24059522658279\\
52 0.180817416011\\
56 0.0907213710746174\\
60 0.941625046119382\\
64 1.18717131206501\\
68 0.447651450200474\\
72 0.00216752049721847\\
76 0.201475553043988\\
80 0.496931432262917\\
84 0.0036358816532176\\
88 0.0989322164843036\\
92 0.568696422258814\\
96 0.58844139136031\\
100 0.000647513015681334\\
104 0.270545429754296\\
108 0.89608842636073\\
112 0.525026891207391\\
116 0.00226322762451736\\
120 0.568254366838911\\
124 1.57862771980131\\
128 1.52623235385953\\
132 0.499684351222474\\
136 0.0021055121953597\\
140 0.489624367063338\\
144 1.14338856233885\\
148 0.550762780600053\\
152 1.67419527403021\\
156 6.39467003439238\\
160 11.3851200070974\\
164 13.2405343618743\\
168 12.1958474948259\\
172 10.7581458496684\\
176 9.26307439345567\\
180 7.83987696110666\\
184 6.52382534228758\\
188 5.27609650672233\\
192 4.10857398492301\\
196 2.92074311831642\\
200 1.62271837935066\\
204 0.218680553670665\\
208 0.130955783129575\\
212 1.08954805861295\\
216 2.11554395383035\\
220 1.21979926575842\\
224 0.109658640471082\\
228 2.49849074360748\\
232 5.79283766335553\\
236 6.31223098828286\\
240 5.14573487480059\\
244 3.79061264768558\\
248 2.54612709348151\\
252 1.37571141397319\\
256 0.237009904995542\\
260 0.0408821525753822\\
264 0.573390425547863\\
268 1.09016032772595\\
272 0.158819528251841\\
276 0.057438039461875\\
280 0.749122728489596\\
284 1.48633863392195\\
288 0.53493577391467\\
292 0.000866123957207754\\
296 0.544926333621586\\
300 1.44041773657207\\
304 0.180276640650955\\
308 3.40542330464128\\
312 7.98639756236645\\
316 9.3659970684203\\
320 8.57695534961133\\
324 7.44484455397193\\
328 6.13865918784594\\
332 4.64593937772645\\
336 3.0458328020046\\
340 1.35421276337172\\
344 0.00723265996319888\\
348 0.514461006217405\\
352 1.90953466033796\\
356 2.7373556235629\\
360 1.70850969220414\\
364 0.300452063250627\\
368 0.0130749525029659\\
372 0.611197785655083\\
376 1.43188707646478\\
380 0.992217272810508\\
384 0.00728152124609949\\
388 0.0927971131165606\\
392 0.60762287300463\\
396 0.761411850081918\\
400 0.00436784927528944\\
};
\end{axis}

\begin{axis}[%
width=0.18\textwidth,
height=0.12\textheight,
scale only axis,
separate axis lines,
every outer x axis line/.append style={darkgray!60!black},
every x tick label/.append style={font=\color{darkgray!60!black}},
xmin=0,
xmax=400,
xlabel={time},
xmajorgrids,
every outer y axis line/.append style={darkgray!60!black},
every y tick label/.append style={font=\color{darkgray!60!black}},
ymin=0,
ymax=9,
ymajorgrids,
name=plot4,
at=(plot2.below south west),
anchor=above north west,
title={$\ell(t)$}
]
\addplot [
color=black,
solid,
line width=1.2,
]
table[row sep=crcr]{
0 0\\
4 1.99619112249754\\
8 3.57479786770082\\
12 1.92282085951797\\
16 0.368637156521911\\
20 0.799254693755631\\
24 2.95910244779927\\
28 3.25801641238556\\
32 1.26083756239975\\
36 0.563736188232338\\
40 2.74349996103217\\
44 3.1368564384419\\
48 1.25717671602018\\
52 0.338592506015308\\
56 2.48288970100632\\
60 2.98515427217841\\
64 1.9813803226033\\
68 0.425832862948385\\
72 1.56747384414152\\
76 2.38573051235797\\
80 2.29258033233274\\
84 1.1243305350589\\
88 2.77117918908952\\
92 2.93538038836825\\
96 1.07451045251403\\
100 1.79435363431592\\
104 2.73132198599334\\
108 2.94409394118515\\
112 0.656801376495839\\
116 1.89703347553781\\
120 2.96767828738535\\
124 2.97426072806622\\
128 0.619201379771926\\
132 0.260701382185784\\
136 1.79726765049158\\
140 2.8357643256118\\
144 2.88944925713828\\
148 0.564760424282539\\
152 4.91477949124408\\
156 7.81573612449777\\
160 8.18753801727652\\
164 1.64386647716184\\
168 0.349217350814789\\
172 0.171212628873204\\
176 0.210056044727471\\
180 0.133373409949117\\
184 0.179803887620811\\
188 0.190708340649013\\
192 0.122954685239372\\
196 0.181598330222899\\
200 0.170478672608544\\
204 0.139308392466004\\
208 3.0259425305484\\
212 3.70476060198573\\
216 2.41137678369215\\
220 0.526094223570131\\
224 4.34906990912486\\
228 6.25916845380066\\
232 5.67461246718144\\
236 0.99083343039181\\
240 0.333858081548792\\
244 0.227435809675859\\
248 0.186858141117418\\
252 0.261451802487981\\
256 0.193025111350243\\
260 2.0884379455491\\
264 2.73297126451231\\
268 1.9391046683401\\
272 0.461187154340327\\
276 2.80307018664246\\
280 3.32130512440982\\
284 2.1072083999545\\
288 0.450337691649016\\
292 2.11581929328254\\
296 3.15060855430177\\
300 2.95869914263276\\
304 5.56309307177816\\
308 7.63903345899347\\
312 7.02586939948634\\
316 1.28330086717065\\
320 0.393582347510232\\
324 0.201215783199233\\
328 0.239583303834656\\
332 0.210560540670223\\
336 0.146051571969931\\
340 0.234809467745537\\
344 0.953240883268952\\
348 3.64827673198374\\
352 4.10865387463147\\
356 1.46651288033863\\
360 0.40380012133579\\
364 0.272324359318804\\
368 2.13393600633416\\
372 2.830654229338\\
376 2.10933754694345\\
380 0.465688627532116\\
384 0.334660201035009\\
388 2.3009483306857\\
392 2.64389312591728\\
396 1.32959961688578\\
400 1.4478501110336\\
};
\end{axis}

\begin{axis}[%
width=0.18\textwidth,
height=0.12\textheight,
scale only axis,
separate axis lines,
every outer x axis line/.append style={darkgray!60!black},
every x tick label/.append style={font=\color{darkgray!60!black}},
xmin=0,
xmax=400,
xlabel={time},
xmajorgrids,
every outer y axis line/.append style={darkgray!60!black},
every y tick label/.append style={font=\color{darkgray!60!black}},
ymin=-2,
ymax=4,
ymajorgrids,
at=(plot4.left of south west),
anchor=right of south east,
title={$u(t)$}
]
\addplot [
color=blue,
dash pattern=on 1pt off 3pt on 3pt off 3pt,
line width=1.2,
]
table[row sep=crcr]{
0 -1\\
4 -1\\
8 -1\\
12 -1\\
16 -1\\
20 -1\\
24 -1\\
28 -1\\
32 -1\\
36 -1\\
40 -1\\
44 -1\\
48 -1\\
52 -1\\
56 -1\\
60 -1\\
64 -1\\
68 -1\\
72 -1\\
76 -1\\
80 -1\\
84 -1\\
88 -1\\
92 -1\\
96 -1\\
100 -1\\
104 -1\\
108 -1\\
112 -1\\
116 -1\\
120 -1\\
124 -1\\
128 -1\\
132 -1\\
136 -1\\
140 -1\\
144 -1\\
148 -1\\
152 -1\\
156 -1\\
160 -1\\
164 -1\\
168 -1\\
172 -1\\
176 -1\\
180 -1\\
184 -1\\
188 -1\\
192 -1\\
196 -1\\
200 -1\\
204 -1\\
208 -1\\
212 -1\\
216 -1\\
220 -1\\
224 -1\\
228 -1\\
232 -1\\
236 -1\\
240 -1\\
244 -1\\
248 -1\\
252 -1\\
256 -1\\
260 -1\\
264 -1\\
268 -1\\
272 -1\\
276 -1\\
280 -1\\
284 -1\\
288 -1\\
292 -1\\
296 -1\\
300 -1\\
304 -1\\
308 -1\\
312 -1\\
316 -1\\
320 -1\\
324 -1\\
328 -1\\
332 -1\\
336 -1\\
340 -1\\
344 -1\\
348 -1\\
352 -1\\
356 -1\\
360 -1\\
364 -1\\
368 -1\\
372 -1\\
376 -1\\
380 -1\\
384 -1\\
388 -1\\
392 -1\\
396 -1\\
400 -1\\
};
\addplot [
color=blue,
dash pattern=on 1pt off 3pt on 3pt off 3pt,
line width=1.2,
]
table[row sep=crcr]{
0 3\\
4 3\\
8 3\\
12 3\\
16 3\\
20 3\\
24 3\\
28 3\\
32 3\\
36 3\\
40 3\\
44 3\\
48 3\\
52 3\\
56 3\\
60 3\\
64 3\\
68 3\\
72 3\\
76 3\\
80 3\\
84 3\\
88 3\\
92 3\\
96 3\\
100 3\\
104 3\\
108 3\\
112 3\\
116 3\\
120 3\\
124 3\\
128 3\\
132 3\\
136 3\\
140 3\\
144 3\\
148 3\\
152 3\\
156 3\\
160 3\\
164 3\\
168 3\\
172 3\\
176 3\\
180 3\\
184 3\\
188 3\\
192 3\\
196 3\\
200 3\\
204 3\\
208 3\\
212 3\\
216 3\\
220 3\\
224 3\\
228 3\\
232 3\\
236 3\\
240 3\\
244 3\\
248 3\\
252 3\\
256 3\\
260 3\\
264 3\\
268 3\\
272 3\\
276 3\\
280 3\\
284 3\\
288 3\\
292 3\\
296 3\\
300 3\\
304 3\\
308 3\\
312 3\\
316 3\\
320 3\\
324 3\\
328 3\\
332 3\\
336 3\\
340 3\\
344 3\\
348 3\\
352 3\\
356 3\\
360 3\\
364 3\\
368 3\\
372 3\\
376 3\\
380 3\\
384 3\\
388 3\\
392 3\\
396 3\\
400 3\\
};
\addplot [
color=black,
solid,
line width=1.2,
]
table[row sep=crcr]{
0 0\\
4 -0.900707916533126\\
8 -1\\
12 -1\\
16 -1\\
20 0.454235113397316\\
24 -1\\
28 -1\\
32 -1\\
36 0.631936253402532\\
40 -1\\
44 -1\\
48 -1\\
52 -1\\
56 -1\\
60 -1\\
64 -1\\
68 -1\\
72 -0.779438734301737\\
76 -1\\
80 -1\\
84 0.152111419962295\\
88 -1\\
92 -1\\
96 -1\\
100 -0.624289874890107\\
104 -1\\
108 -1\\
112 -1\\
116 -0.952436752489988\\
120 -1\\
124 -1\\
128 -1\\
132 -1\\
136 -0.987485078306002\\
140 -1\\
144 -1\\
148 -1\\
152 -1\\
156 -1\\
160 -1\\
164 -1\\
168 -1\\
172 -1\\
176 -1\\
180 -1\\
184 -1\\
188 -1\\
192 -1\\
196 -1\\
200 -1\\
204 -1\\
208 -1\\
212 -1\\
216 -1\\
220 -1\\
224 -1\\
228 -1\\
232 -1\\
236 -1\\
240 -1\\
244 -1\\
248 -1\\
252 -1\\
256 -1\\
260 -1\\
264 -1\\
268 -1\\
272 -1\\
276 -1\\
280 -1\\
284 -1\\
288 -1\\
292 -0.989265911574334\\
296 -1\\
300 -1\\
304 -1\\
308 -1\\
312 -1\\
316 -1\\
320 -1\\
324 -1\\
328 -1\\
332 -1\\
336 -1\\
340 -1\\
344 0.361723894136061\\
348 -1\\
352 -1\\
356 -1\\
360 -1\\
364 -1\\
368 -1\\
372 -1\\
376 -1\\
380 -1\\
384 0.35142864756151\\
388 -1\\
392 -1\\
396 -1\\
400 -0.130430762439109\\
};
\end{axis}
\end{tikzpicture}%